\documentclass[a4paper]{revtex4}
\usepackage{graphicx}
\usepackage{amssymb}
\usepackage{color}
\usepackage{ulem}
\usepackage{dcolumn}
\usepackage{bm}
\begin{document}

~
\setcounter{page}{0}
\thispagestyle{empty}
\vspace{20cm}

Publishing information:

J.V.I. Timonen, R.H.A. Ras, O. Ikkala, M. Oksanen, E. Sepp\"al\"a, K. Chalapat, J. Li, G.S. Paraoanu, {\it Magnetic nanocomposites at microwave frequencies}, in {\sl Trends in nanophysics: theory, experiment, technology}, edited by V. Barsan and A. Aldea,  Engineering Materials Series, Springer-Verlag, Berlin (ISBN: 978-3-642-12069-5), pp. 257-285 (2010).

DOI: 10.1007/978-3-642-12070-1\_11

\newpage

\title{Magnetic nanocomposites at microwave frequencies}

\author{Jaakko V. I. Timonen}
\author{Robin H. A. Ras}
\author{Olli Ikkala}
\affiliation{Molecular Materials, Department of Applied Physics, School of Science and Technology, Aalto University, P. O. Box 15100, FI-00076 AALTO, Finland.}

\author{Markku Oksanen}
\author{Eira Sepp\"al\"a}
\affiliation{Nokia Research Center, It\"amerenkatu 11-13, 00180 Helsinki, Finland.}

\author{Khattiya Chalapat}
\author{Jian Li}
\author{Gheorghe Sorin Paraoanu}
\affiliation{Low Temperature Laboratory, School of Science and Technology, Aalto University, P. O. Box 15100, FI-00076 AALTO, Finland.}

\date{\today}

\begin{abstract}
Most conventional magnetic materials used in the electronic devices are ferrites, which are composed of micrometer-size grains. But ferrites have small saturation magnetization, therefore the performance at GHz frequencies is rather poor. That is why functionalized nanocomposites comprising magnetic nanoparticles ({\it e.g.} Fe, Co) with dimensions ranging from a few nm to 100 nm, and embedded in dielectric matrices ({\it e.g.} silicon oxide, aluminium oxide) have a significant potential for the electronics industry. When the  size of the nanoparticles is smaller than the critical size for multidomain formation, these nanocomposites can be regarded as an ensemble of particles in single-domain states and the losses (due for example to eddy currents) are expected to be relatively small.

Here we review the theory of magnetism in such materials, and we present a novel measurement method used for the characterization of the electromagnetic properties of composites with nanomagnetic insertions. We also present a few experimental results obtained on composites consisting of iron nanoparticles in a dielectric matrix.

\end{abstract}

\maketitle

\section{Introduction}
\label{intro}

For a long time have ferrites been the best choice of material for various applications requiring magnetic response at radio frequencies (RF). In recent times, there has been a strong demand both from the developers and the end-users side for decreasing the size of the modern-day portable communication devices and to add new functionalities that require access to broader communication bands or to other bands than those commonly used in communication between such devices. All this should be achieved without increasing power consumption; rather, a decrease would be desired. The antenna for example is a relatively large component of modern-day communication devices. If the size of the antenna is decreased by a certain factor, then the resonance frequency of the antenna is increased by the same factor \cite{shirakata}. As a result, in order to compensate this increase in the resonance frequency, the antenna cavity may be filled with a material in which the wavelength of the external radiation field is reduced by the same factor. The wavelength $\lambda$ inside a material of relative dielectric permittivity $\epsilon$ and the relative magnetic permeability $\mu$ is given by $\lambda = \lambda_{0}/\sqrt{\epsilon\mu}$
where $\lambda_{0}$ is the wavelength in vacuum. Hence, it is possible to decrease the wavelength inside the antenna - and therefore also the size of the antenna - by increasing the permittivity or the permeability or the both. Once the size reduction is fixed - that is, $\epsilon\mu$ is fixed - the relative strength between the permittivity and the permeability needs to be decided. It is known that the balance between these two affects the bandwidth of the antenna. Generally speaking, high-$\epsilon$ and low-$\mu$ materials decrease the bandwidth of the microstrip antenna while low-$\epsilon$ and high-$\mu$  materials keep the bandwidth unchanged or even increase it \cite{hansen}.

Typical high-$\mu$ materials are magnetically soft metals, alloys, and oxides. Of these, metals and alloys are unsuitable for high-frequency applications since they are conducting. On the other hand, non-conducting oxides - such as the ferrites mentioned above - have been used and are still being used in many applications. Their usefulness originates from poor conductivity and the ferrimagnetic ordering. But ferrites are limited by low saturation magnetization which results in a low ferromagnetic resonance frequency and a cut-off in permeability below the communication frequencies \cite{waldron}. The ferromagnetic resonance frequency has to be well above the designed operation frequency to avoid losses and to have significant magnetic response. However, modern standards such as the Global System for Mobile communications (GSM), the Wireless Local Area Network (WLAN), and the Wireless Universal Serial Bus (Wireless USB) operate in the Super High Frequency (SHF) band or in its immediate vicinity \cite{super}. The frequency range covered by the SHF band is 3-30 GHz and it cannot be accessed by the ferrites whose resonance frequency is typically of the order of hundreds of MHz \cite{waldron}. Hence, other kinds of materials need to be developed for the applications mentioned.

The important issue related to the miniaturization by increasing the permittivity and/or the permeability is the introduced energy dissipation. In some contexts, losses are good in a sense that they reduce the resonance quality factor and hence increase the bandwidth. The cost is increased energy consumption which goes to heating of the antenna cavity.
In general, several processes contribute to losses in magnetic materials. At low frequencies, the dominant loss process is due to hysteresis: it becomes less important as the frequency increases, due to the fact that the motion of the domain walls becomes dampened. The eddy current loss plays a dominant role in the higher-frequency range: the power dissipated in this process scales quadratically with frequency. In this paper will have a closer look at this source of dissipation, which can be reduced in principle by using nanoparticles instead of bulk materials. Another important process which we will discuss is ferromagnetic resonance (due to rotation of the magnetization).

All these phenomena limit the applicability of standard materials for high-frequency electronics. However, the SHF band may be accessed by the so called magnetic granularmaterials. A granular material is composed of a non-conducting matrix with small (metallic) magnetically soft inclusions. Such composites have both desired properties; they are non-conducting and magnetically soft. Granular materials are of special interest at the moment since the synthesis of extremely small magnetic nanoparticles has taken major leaps during the past decades. Especially the synthesis of monodisperse FePt nanoparticles \cite{sun} and the synthesis of shape and size controlled cobalt nanoparticles \cite{puntes} have generated interest because these particles can be produced with a narrow size distribution. In addition, small nanoparticles exhibit an intriguing magnetic phenomenon called superparamagnetism. Superparamagnetic nanoparticles are characterized by zero coercivity and zero remanence which can lead to a decrease in loss in the magnetization process \cite{moerup}.

There have been numerous studies investigating dielectric and magnetic responses of different
granular materials. For example, an epoxy-based composite containing 20\% (all percentages in this article are defined as volume per volume) rod-shaped CrO$_2$ nanoparticles has been demonstrated to have a ferromagnetic resonance around 8 GHz and relative permeability of 1.2 \cite{wu}. Similarly, a multimillimetre-large self-assembled superlattice of 15 nm FeCo nanoparticles has been shown  to have a ferromagnetic resonance above 4 GHz \cite{desvaux}.

This raises the interesting question of whether it would be possible in general to design novel nanocomposite materials with specified RF and microwave electromagnetic properties, aiming for example at very large magnetic permeabilities and low loss at microwave frequencies. Such properties should arise from the interparticle exchange coupling effects which, for small enough interparticle separation, extends over near-neighbour particles, and from the reduction of the eddy currents associated with the lower dimensionality of the particles. In this paper, we aim at evaluating the feasibility of using magnetic polymer nanocomposites as magnetically active materials in the SHF band.

The  structure of the paper is the following: in Section \ref{magnetism} we review briefly the physics of ferromagnetism in nanoparticlee, namely the existence of single-domain states (Subsection \ref{single}), ferromagnetic resonance and the Snoek limit (Subsection \ref{eddy}), and eddy currents (Subsection \ref{eddy}). In Section \ref{nanocomposites} we discuss theoretically issues such as the requirements stated by thermodynamics on the possibility of dispersing nanoparticles in polymers (Subsection \ref{disp}). A set of rules governing the effective high-frequency magnetic response in magnetic nanocomposites is developed in Subsection \ref{rules}. Then we describe the experimental details and procedures used to prepare and characterize the nanocomposites (Section \ref{ch}). We continue to Section \ref{hf} where we first discuss a measurement protocol which allow us to measure the electromagnetic properties of the iron nanocomposites (Subsection \ref{prop}).
Finally, as the main experimental result of this paper, magnetic permeability and dielectric permittivity spectra between 1-14 GHz are reported in Subsection \ref{ss} for iron-based nanocomposites (containing Fe/FeO nanoparticles in a polystyrene matrix) as a function of the nanoparticle volume fraction.
This paper ends with a discussion (Section \ref{conc}) on how to improve the magnetic performance in the SHF band.

\section{Magnetism in nanoparticles}
\label{magnetism}

Magnetic behavior in ferromagnetic nanoparticles is briefly reviewed in this section {\it c.f.} \cite{chikazumi}-\cite{kittel}. The focus is especially in the so called single-domain magnetic nanoparticles which lack the typical multi-domain structure observed in bulk ferromagnetic materials. The topics to be discussed are: A) when does the single-domain state appear, B) what is its ferromagnetic resonance frequency, and C) what are the sources of energy dissipation in single-domain nanoparticles.

\subsection{Existence criteria for the single-domain state}
\label{single}

A magnetic domain is a uniformly magnetized  region within a piece of ferromagnetic or ferrimagnetic material. Magnetic domains are separated by boundary regions called the domain walls (${\rm DW}$) in which the magnetization gradually rotates from the direction defined by one of the domains to the direction defined by the other. The domain wall thickness ($d_{\rm DW}$), which depends on the material's exchange stiffness coefficient ($A$) and the anisotropy energy density  ($K$), extends from 10 nm in high-anisotropy materials to 200 nm in low-anisotropy materials.
The domain thickness, on the other hand, depends more on geometrical considerations. For example, in one square centimeter iron ribbon, 10 $\mu$m thick, the domain wall spacing is of the order of 100 $\mu$m. The spacing increases if the thickness is reduced. Reducing the thickness over a critical value leads to the complete disappearance of the domain walls. That state is called the single-domain (${\rm SD}$) state. Between multidomain and single-domain states there may be a vortex state: this is not discussed however here. Similarly, the domains in spherical nanoparticles vanish below a certain diameter which is of the order of few nanometers or few tens of nanometers. In hard materials this diameter ($d_{\rm SD,HARD}$) can be estimated to be roughly (\cite{handley}, p. 303):
\begin{equation}
d_{\rm SD, HARD} \approx 18 \frac{\sqrt{AK}}{\mu_{0}M_{\rm S}^{2}},\label{hard}
\end{equation}
where $M_{\rm S}$ is the saturation magnetization and $\mu_{0}$ is the vacuum permeability. The equation is based on the assumption that the magnetization follows the energetically favorable directions (easy axes or easy planes) defined by the anisotropy. The single-domain diameter given by Eq. (\ref{hard}) should be always compared to the domain wall thickness given by (\cite{handley}, p. 283)
\begin{equation}
d_{\rm DW}=\pi\sqrt{\frac{A}{K}}.\label{dw}
\end{equation}
If the diameter of the particle is less than the wall thickness, it is obvious that it cannot support the wall. The condition $d_{\rm SD,HARD}>d_{\rm DW}$, leads to the criterion
\begin{equation}
Q \stackrel{\rm def}{=} \frac{18}{\pi}\frac{K}{\mu_{0}M_{\rm S}^2} >1 . \label{criterion}
\end{equation}
On the other hand, in magnetically soft nanoparticles the magnetization does not necessary follow the easy directions. In the perfectly isotropic case, that is $K=0$, the surface spins are oriented along the spherical surface and a vortex core is formed in the center of the particle if the particle is above the single-domain limit. The single-domain diameter ($d_{\rm SD,SOFT}$) of a perfectly isotropic nanoparticle is given by (\cite{handley}, p. 305),
\begin{equation}
d_{\rm SD, SOFT}\approx 6\sqrt{\frac{A}{\mu_{0}M_{\rm S}^2}\left[\ln\frac{d_{\rm SD,SOFT}}{a}-1\right]},\label{soft}
\end{equation}
where $a$ is the lattice constant. This equation can be solved by the iteration method. The single-domain diameter is more difficult to estimate if the anisotropy is non-zero but does not meet the requirement of Eq. (\ref{criterion}). In that case, the single-domain diameter is likely to rest between the values predicted by Eqs. (\ref{hard}) and (\ref{soft}).

Single-domain diameters, domain wall thicknesses and other relevant physical quantities for selected ferromagnetic metals are shown in Table \ref{table1}. The additional surface-induced anisotropy has been neglected. The uniaxial hexagonal close packed (HCP) cobalt is the only strongly anisotropic material with $Q\approx 1$. The body centered cubic (BCC) iron and the FCC nickel fall in between hard and soft behavior.

\begin{table}
\center
\caption{The saturation magnetization ($M_{\rm S}$)  \cite{sorensen}, the anisotropy energy density $K$ \cite{sorensen} \cite{steinmuller}, the Q-factor Eq. (\ref{criterion}), the single-domain diameter in the hard material approximation $d_{\rm SD,HARD}$ Eq. (\ref{hard}), single-domain diameter in the isotropic material limit $d_{\rm SD,SOFT}$ Eq. (\ref{soft}), and the domain wall width $d_{\rm DW}$ Eq. (\ref{dw}), for iron, cobalt, and nickel. The exchange stiffnesses used in the calculations are from \cite{george}.}

\begin{tabular}{||c|c|c|c|c|c|c||}\hline\hline
    & $M_{\rm S}$ & $K$ & $Q$ & $d_{\rm SD,HARD}$ & $d_{\rm SD,SOFT}$ & $d_{\rm DW}$ \\
   & (emu/cm$^3$) & (erg/cm$^3$) &  & (nm) & (nm) & (nm) \\
    \hline
   &  &  &  & & &  \\
Iron (BCC) & 1707 & $4.8\times 10^5$ &  0.075 & 5 & 89 & 63 \\
\hline
 &  &  &  & & &  \\
Cobalt (HCP) & 1440 & $4.5\times 10^6$ &  0.996 & 26 & 169 & 26 \\
\hline
&  &  &  & & &  \\
Nickel (FCC) & 485 & $-5.7\times 10^4$ &  0.110 & 13 & 173 & 113 \\
\hline
\hline

\end{tabular}
\label{table1}
\end{table}

\subsection{Ferromagnetic resonance and the Snoek limit}
\label{snoek}

The two major processes contributing to the magnetization change are the domain wall motion and the domain rotation. The resonance frequency of the domain wall motion is typically less than the resonance frequency of the domain rotation. Hence, the only process active in the highest frequencies is the domain rotation which is associated with the ferromagnetic resonance (FMR).

The natural \footnote{"Natural" is used so that this resonance is distinguished from the dimensional resonance associated with standing waves within a sample of finite size. The dimensional resonance takes place when the end-to-end length ($L$) of the sample times two is equal to an integer multiple of the wavelength of the radiation ($\lambda$) in the material. The same is mathematically expressed as $L=n\lambda /2=n/2 \lambda_{0}/\sqrt{\epsilon\mu}$ where $\lambda_{0}$ is the wavelength in vacuum and $n$ is a positive integer. The resonance frequency ($f_r$) is given by $f_{r}=c/\lambda_{0} =nc/2L\sqrt{\epsilon\mu}$ where $c$ is the speed of light. For example, the first dimensional resonance in a 10 mm long sample with $\epsilon\mu=9$ is approximately 5 GHz. The dimensional resonance can be avoided in experiments by carefully estimating the product $\epsilon\mu$ and designing the sample length accordingly.} ferromagnetic resonance was first explained by Snoek to be the resonance of the magnetization vector ($\vec{M}$) pivoting under the action of some energy anisotropy field ($\vec{H}_{\rm A}$) \cite{snoek}. The origin of the anisotropy is not restricted. It can be induced, for example, by an external magnetic field, magnetocrystalline anisotropy or shape anisotropy. It is common to treat any energy anisotropy as if it was due to an external magnetic field.

The motion of the magnetization around in the anisotropy field is described by the Landau-Lifshitz equation \cite{landau},
\begin{equation}
\frac{d\vec{M}}{dt}=-\nu (\vec{M}\times \vec{H}_{\rm A})-\frac{4\pi\mu_{0}\hat{\lambda}}{M_{\rm S}^2}(\vec{M}\times (\vec{M}\times\vec{H}_{\rm A})),
\end{equation}
where $\hat{\lambda}$ is the relaxation frequency (not the resonance frequency) and $\nu$  is the gyromagnetic constant given by (\cite{chikazumi}, p. 559)
\begin{equation}
\nu =g \frac{e\mu_{0}}{2m}\approx 1.105 \times 10^{5} g ({\rm mA^{-1}s^{-1}}) \approx 2.2 \times 10^{5} {\rm mA^{-1}s^{-1}}, \label{nu}
\end{equation}
where $g$ is the gyromagnetic factor (taken to be 2), $e$ is the magnitude of the electron charge and $m$ is the electron mass.

If the Landau-Lifshitz equation is solved, one obtains the resonance condition (\cite{chikazumi} p. 559)
\begin{equation}
f_{{\rm FMR}}=(2\pi )^{-1} \nu H_{{\rm A}}, \label{ll}
\end{equation}
where $f_{{\rm FMR}}$ is the resonance frequency and $H_{{\rm A}}$ is the magnitude of the anisotropy field.

For example, for HCP cobalt the magnetocrystalline anisotropy energy density $\left(U_{{\rm A}}\right)$ is given by (\cite{chikazumi}, p. 264)
\begin{equation}
U_{{\rm A}}=K{{\sin }^2 \theta \ }\approx K\left({\theta }^2-\frac{1}{3}{\theta }^4+\dots \right),
\end{equation}
where $\theta $ is the angle between the easy axis and the magnetization. The energy density due to an imaginary magnetic field is given by (\cite{chikazumi}, p. 264)
\begin{equation}
U_{{\rm A}}=-{\mu }_0H_{{\rm A}}M_{{\rm S}}{\cos  \theta \ }\approx -{\mu }_0H_{{\rm A}}M_{{\rm S}}\left(1-\frac{1}{2}{\theta }^2+\dots \right).
\end{equation}
By comparing the exponents one obtains
\begin{equation}
H_{{\rm A}}=\frac{2K}{{\mu }_0M_{{\rm S}}}\approx \frac{0.62{\rm \ T}}{{\mu }_0},\label{unspe}
\end{equation}
and from Eq. (\ref{ll})
\begin{equation}
f_{{\rm FMR}}=(2\pi )^{-1} \nu H_{{\rm A}}\approx 17{\rm \ GHz}.
\end{equation}

It is tempting to use nanoparticles with as high anisotropy as possible in order to maximize the FMR frequency. Unfortunately, the permeability decreases with the increasing anisotropy; for uniaxial materials the relative permeability $\mu$ is given by (\cite{chikazumi}, p. 493),
\begin{equation}
\mu=1+\frac{\mu_{0}M^2_{{\rm S}}{{\sin }^{{\rm 2}} \theta \ }}{2K}.\label{treispe}
\end{equation}
It is easy to show that that Eqs. (\ref{ll}),(\ref{unspe}), and (\ref{treispe}) lead to
\begin{equation}
\left\langle \mu \right\rangle \cdot f_{{\rm FMR}} =\frac{\nu M_{{\rm S}}}{3\pi},\label{result}
\end{equation}
where $\left\langle \mu \right\rangle $ is the angular average of the relative permeability (which we assume much larger than the unit). This equation is known as the Snoek limit. It is an extremely important result since it predicts the maximum permeability achievable with a given FMR frequency as a function of the saturation magnetization. It can be shown to be valid for both the uniaxial and cubic materials (taken that$\ K>0$). Some values for the maximum relative permeability as a function of the FMR frequency and the saturation magnetization are shown in Table \ref{table2}.

It has been found out that the Snoek limit can be exceeded in materials of negative uniaxial anisotropy \cite{jonker}. In that case, the magnetization can rotate in the easy plane perpendicular to the c-axis. Such materials obey the modified Snoek limit (\cite{chikazumi}, p. 561)
\begin{equation}
\mu \cdot f_{{\rm FMR}} =\frac{\nu M_{{\rm S}}}{3\pi}\sqrt{\frac{H_{{\rm A1}}}{H_{{\rm A2}}}},
\end{equation}
where $H_{{\rm A}1}$ is the anisotropy field along the c-plane (small) and $H_{{\rm A}2}$ is the anisotropy field out of the c-plane (large). One such material is the Ferroxplana \cite{kittel}.

\begin{table}
\center
\caption{Maximum relative permeability $\left(\mu \right)$ Eq. (\ref{result}) achievable in cubic and uniaxial materials with positive anisotropy as a function of the saturation magnetization $\left(M_{{\rm S}}\right)$ and the FMR frequency $\left(f_{{\rm FMR}}\right)$.}

\label{tabel2}
\begin{tabular}{c||c|c|c|c|c}\hline\hline
 &  \multicolumn{5}{c}{Saturation magnetization $\mu_{0}M_{\rm S}$}\\
   \hline
   $f_{\rm FMR}$ &  &  &   &   \\
 (GHz)    & 0.1 T & 0.3 T & 0.5 T & 1.0 T & 2.0 T\\
    \hline \hline
   &  &  &  & &  \\
0.1 & 19.7 & 57.7 & 94.3  &  187.60 & 374.2  \\
\hline
 &  &  &  & &   \\
0.5 & 4.7 & 12.2 &  19.7 & 38.3 & 75.6  \\
\hline
&  &  &  & &   \\
1.0  & 2.9 & 6.6 &  10.3 & 19.7 & 38.3  \\
\hline
&  &  &  & &   \\
2.0 & 1.9  & 3.8 & 5.7 &  10.3 & 19.7 \\
\hline
&  &  &  & &  \\
5.0  & 1.4 & 2.1 &  2.9 & 4.7 & 8.5 \\
\hline
\hline

\end{tabular}
\label{table2}
\end{table}

\subsection{Eddy currents and other sources of loss}
\label{eddy}

Magnetic materials can dissipate energy through various processes when magnetized. When the oscillation period of the external driving field is long, the main sources of loss are the processes that contribute to the hysteresis. The hysteresis loss is linearly proportional to the frequency of the driving field since the loss during one complete hysteresis cycle (\textit{B-H} loop) is proportional to the area within the cycle (assuming that the hysteresis loop does not change with the frequency). The main contribution to the hysteresis comes from the domain wall motion and pinning and a smaller contribution is due to the magnetization rotation and domain nucleation. The domain wall motion is damped as the frequency is increased over the domain wall resonance so that only the magnetization rotation persists to the highest frequencies. In addition to the domain rotation hysteresis, the loss in the SHF band stems also from the electrical currents induced by the changing magnetic field inside the particles

A change in the magnetic field $\left(B\right)$ inside a piece of material with finite resistivity $\left(\rho \right)$ induces an electric field which generates an electric current as stated by the Faraday's law. This current is called eddy current. It dissipates energy into the sample through the electrical resistance. For example, the averaged loss power $\left\langle P\right\rangle $ in a spherical nanoparticle of radius $r$ can be calculated to be \footnote{Assume that a spherical nanoparticle (radius $r$, resistivity $\rho $ ), is placed in an alternating magnetic field so that the magnetic field inside the particle is $B$. According to the Faraday's equation  $2\pi xE\left(x\right)=-\pi x^2 dB/dt$  where  $x$ is the distance from the particle center and  $E\left(x\right)$  is the electric field. The differential current $dI$ circulating around the cylindrical shell at the distance $x$ is given by  $dI=2\pi xE\left(x\right) /dR$  where the differential resistance $dR$ is given by  $dR=\rho 2\pi x /h\left(x\right)dx$  where  $h(x)$  is the height of the cylindrical shell and $dx$ is the thickness of the shell. Now the dissipated power can be calculated from $P=\int{2\pi xE\left(x\right)dI}=(\pi /\rho ){\left(dB/dt\right)}^2\int^r_0{x^3\sqrt{r^2-x^2}dx}$. Changing the integration variable from $x$ to $r{\sin  \varphi }$  simplifies the integral to  $P= (\pi /\rho )\left(dB/dt\right)^2r^5\int^{\pi / 2}_0{\left({{\sin }^{{\rm 3}} \varphi }-{{\sin }^{{\rm 5}} \varphi }\right)d\varphi
}$  where the integral part is equal to 2/15. Assuming that the magnetic field is given by  $B=\hat{B}{\sin  2\pi f\ }$  leads to the result given in Eq. (\ref{loss}).}
\begin{equation}
\left\langle P\right\rangle =\frac{2\pi }{15}\frac{1}{\rho }\ r^5\left\langle {\left(\frac{dB}{dt}\right)}^2\right\rangle =\frac{4{\pi }^3}{15\rho }r^5{\left(f\hat{B}\right)}^2 ,
\label{loss}
\end{equation}
where $f$ is the frequency of the driving field and $\hat{B}$ is the amplitude  of the  oscillating component of the total magnetization. From Eq. (\ref{loss}) it is obvious that the loss power per unit volume increases as
$r^2$, indicating that the loss can be decreased by using finer nanoparticles. Notice that the loss power will vanish above the FMR resonance since there cannot be magnetic response above that frequency, that is $\hat{B}\to 0$.

For example, the loss power per unit volume $\left(p\right)$ in cobalt nanoparticles can be calculated to be
\begin{equation}
p\approx {\rm 32}{\left(\frac{r}{{\rm nm}}\right)}^{{\rm 2}}{\left(\frac{f}{{\rm GHz}}\right)}^{{\rm 2}}{\left(\frac{\hat{B}}{{\rm T}}\right)}^{{\rm 2}}\frac{{\rm W}}{{\rm c}{{\rm m}}^{{\rm 3}}} .
\end{equation}
If the volume of the magnetic element is$\ {\rm 0.1\ c}{{\rm m}}^{{\rm 3}}$, the radius of the nanoparticles 5 nm, and $\hat{B}=1.8$ T one obtains 0.26 mW for loss power.

It has been shown that this simple approach is inadequate to describe the eddy current loss in materials containing domain walls \cite{chikazumi}. The eddy currents in multidomain materials are localized at the domain walls, which leads to a roughly four-times increase in the loss. However, since there are no walls present in single-domain nanoparticles and the magnetization reversal can take place by uniform rotation, this model is considered here to be adequate in describing the eddy current loss in single-domain nanoparticles.

\noindent One more matter to be addressed is the penetration depth of the magnetic field into the nanoparticles. Because the eddy currents create a magnetic field counteracting the magnetic field that induced the eddy currents, the total magnetic field is reduced when moving from the nanoparticle surface towards its core. The depth $\left(s\right)$ at which the magnetic field is reduced by the factor 1/\textit{e} is called the skin-depth and it is given by (\cite{chikazumi}, p. 552),
\begin{equation}
s=\sqrt{\frac{2\rho }{\omega \mu \mu_{0}}}
\label{an}.
\end{equation}
For example, from Eq. (\ref{an}) the skin-depth for cobalt ($\rho {\rm =62\ n}\Omega {\rm m}$ and $\mu {\rm =10}$) at 1 GHz is 1.3 $\mu$m and at 10 GHz 400 nm. Hence, cobalt nanoparticles that are less than 100 nm in diameter would already be on the safe side. The situation is rather different in typical ferrites for which $\rho \approx {{\rm 10}}^{{\rm 4}}{\rm \ }\Omega {\rm m}$ and ${\rm \ \ }\mu {\rm =}{{\rm 10}}^{{\rm 3}}$, giving 5 cm for the skin depth. Therefore  ferrites can be used in the bulk form in near-microwave applications.

\section{MAGNETIC POLYMER NANOCOMPOSITES}
\label{nanocomposites}

\begin{figure}[htb]
\includegraphics[width=9cm]{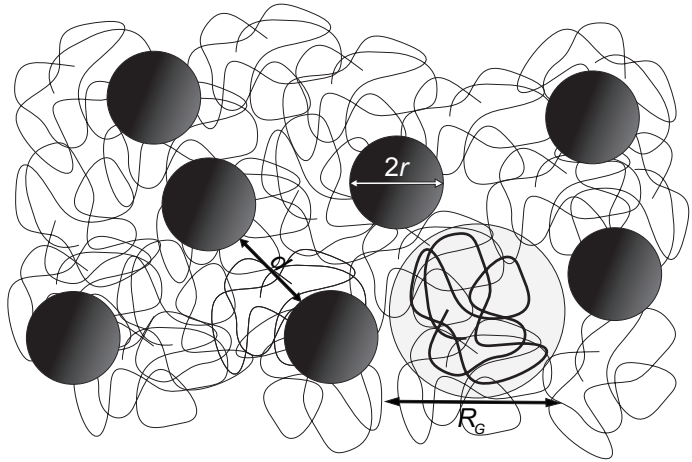}
\caption{A schematic illustration of a polymer nanocomposite. The average radius of the nanoparticles $\left(2r\right)$\textit{ }(filled dark circles), the average radius of gyration of the polymer molecules $\left(R_{{\rm G}}\right)$ (the thick black line inside the filled light-gray circle) and the average nearest-neighbor distance $\left(d\right)$ between the nanoparticles are of the same magnitude.}
\label{polymers}
\end{figure}

In the simplest form a polymer nanocomposite is a blend of small particles (the diameter is less than 100 nm) incorporated in a polymeric matrix. Polymer nanocomposites are characterized by the convergence of three different length scales: the average radius of gyration of the polymer molecules $\left(R_{{\rm G}}\right)$, the average diameter of the nanoparticles $\left(2r\right)$, and the average nearest-neighbor distance between the particles $\left(d\right)$,  as shown in Fig. \ref{polymers}. In such composites, the polymer chains may not adopt bulk-like conformations \cite{krishnamoorti}. Associated with this, there can be a change in the polymer dynamics which can lead to either an increase or a decrease in the glass transition temperature. Furthermore, the nanoparticles bring their own flavor to the nanocomposite - magnetism, in our particular case.

The most severe problem faced in polymer nanocomposites is the aggregation of nanoparticles. The thermodynamic stability of the nanoparticle dispersion has been addressed in the recent literature experimentally, theoretically and through computer simulations. The experiments have showed that nanoparticles aggregate even at small particle volume fractions -- less than 1\% in many compositions \cite{mackay}. Theoretical considerations and computer simulations have revealed that the quality of the nanoparticle dispersion depends delicately on the balance between the entropic and the enthalpic contributions -- quite similarly as in polymer blends \cite{balazs}. The solution for the dispersion dilemma has been pursued by modifying the nanoparticle surface, changing the architecture and size of the polymer and by applying alternative processing conditions.

The simulation results and the theoretical arguments presented in the literature are often difficult to interpret. Furthermore, they do not take into account the magnetic interactions in magnetic nanocomposites. The aim of subsection \ref{disp} is to analyze the factors affecting the dispersion quality of magnetic nanoparticles in non-magnetic polymers. Subsection \ref{rules}  discusses the effective magnetic response of such nanocomposites.

\subsection{Factors Affecting the Nanoparticle Dispersion Quality}
\label{disp}

\subsubsection{Attractive Interparticle Interactions}

There has been considerable interest in modifying chemically the nanoparticle surface towards being more compatible with the polymer \cite{glogowski}, \cite{latham}. Especially important surface modification techniques are the grafting-techniques. They involve either a synthesis of polymer molecules onto nanoparticle surface (grafting-from) or attachment of functionalized polymers onto the the nanoparticle surface (grafting-to). The advantage of the grafting-techniques is that they can make the nanoparticle surface not only enthalpically compatible with a polymer, but the grafted chains also exhibit similar entropic behavior as the surrounding polymer molecules. One disadvantage is that these techniques require precise knowledge of the chemistry involved.

It is well-established that a monolayer of small molecules attached to the nanoparticle surface is not enough to significantly enhance the quality of the dispersion even if the surface molecules were perfectly compatible with the polymer -- that is, they were identical to the constitutional units of the polymer. This is due to the fact that the London dispersion force \footnote{ Despite of its name it is an attractive force.} acting between the nanoparticles is effective over a length which increases linearly with the nanoparticle diameter. This is proven in the following.

\begin{figure}[htb]
\includegraphics[width=12cm]{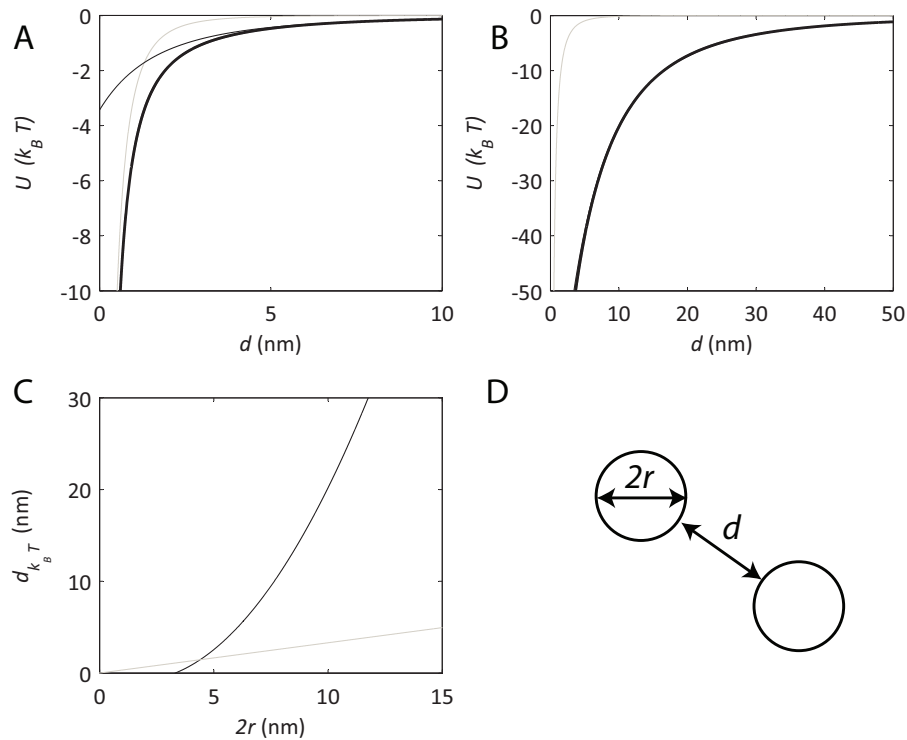}
\caption{Comparison between the London dispersion force and the magnetic dipolar interaction between two identical metal nanoparticles. A) The reduced London potential Eq. (\ref{london}) (grey thin curve), the magnetic dipolar energy Eq. (\ref{26}) (black thin curve) and the total interaction energy (black thick curve) between two 5 nm metal nanoparticles. B) The same for two metal particles 15 nm in diameter. The magnetic dipolar energy curve is overlapping with the total interaction curve. C) The distance between the particle surfaces as a function of the particle diameter when the interaction energy is comparable to the thermal energy. The black line corresponds to the magnetic interaction Eq. (\ref{27}) and the grey to the London dispersion Eq. (\ref{21}). D) Schematic illustration and definition of the used variables.}
\label{figure2}
\end{figure}

The London dispersion energy $\left(U_{{\rm LONDON}}\right)$ between two identical spheres, diameters $2r$, separated by a distance \textit{d} was first shown by Hamaker to be \cite{isra},
\begin{equation}
U_{{\rm LONDON}}=-\frac{A_{121}}{6}\left[\frac{{\left(2r\right)}^2}{2{\left(2r+d\right)}^2}+\frac{{\left(2r\right)}^2}{2\left({\left(2r+d\right)}^2-{\left(2r\right)}^2\right)}+{\ln  \left(1-\frac{{\left(2r\right)}^2}{{\left(2r+d\right)}^2}\right)\ }\right], \label{london}
\end{equation}
where $A_{121}$ is the effective Hamaker for the nanoparticles (phase 1) immersed in the polymeric matrix (phase 2). The Hamaker constants are typically listed for two objects of the same material in vacuum from which the effective value can be calculated by using the approximation \cite{isra}
\begin{equation}
A_{121}\approx {\left(\sqrt{A_{11}}-\sqrt{A_{22}}\right)}^2,\label{constant}
\end{equation}
where $A_{11}$ is the Hamaker constant for the nanoparticles and ${{\rm A}}_{{\rm 22}}$ is the Hamaker constant for the medium. The typical effective Hamaker constant for metal particles immersed in organic solvent or a polymer is approximately$\ 25\cdot {10}^{-20}{\rm \ J}$. By using this value, the London potential Eq. (\ref{london}) is plotted for 5 nm metal particles in Figure \ref{figure2}A and for 15 nm particles in Figure \ref{figure2}B. The distance $\left(d_{k_{{\rm B}}T{\rm ,LONDON}}\right)$ over which the London dispersion force is effective can be estimated by setting the interaction energy equal to the thermal energy and by solving for the distance. The result is \footnote{By setting  $U_{{\rm LONDON}}={-k}_{{\rm B}}T$  and defining a reduced variable  $\alpha =1+ d_{k_{{\rm B}}T,{\rm LONDON}}/2r$  one can rewrite Eq. (\ref{london}) as $6k_{{\rm B}}T/A_{121}=1/(2{\alpha }^2)+ 1/[2\left({\alpha }^2-1\right)]+{\ln  \left(1- 1/{\alpha }^2\right)\ }\equiv f\left(\alpha \right)$ . This equation can be solved for  $\alpha $  by plotting  $y=f\left(\alpha \right)$  and  $y=6k_{{\rm B}}T/A_{121}$  and by locating the point of intersection. The effective distance can be calculated by inserting the obtained intersection point into the equation defining the reduced variable.}
\begin{equation}
d_{k_{{\rm B}}T{\rm ,LONDON}}=\left(\alpha -1\right)\cdot 2r\approx \frac{2r}{3},\label{21}
\end{equation}
where $\alpha $ is a constant in excess of unity and typically around 1.33 for metals immersed in organic medium. This linear dependence is shown in Figure \ref{figure2}C. Typically, nanoparticles are covered with a monolayer of alkyl chains ranging up to 20 carbon-carbon bonds in length. Even if the chains were totally extended and rigid, their length would be only roughly 2 nm. Such a shielding layer can protect only nanoparticles less than 12 nm in diameter from aggregation.

Fortunately, the thermodynamic equilibrium is not solely dependent on the enthalpy which always drives the system towards the phase separation. The additional component is entropy which opposes the separation. The Gibbs free energy\textit{ }(\textit{G}) which determines the thermodynamic stability in the constant temperature and the constant pressure is given by $G = H  -  TS$, where \textit{H} is enthalpy and \textit{S }is entropy. The entropic term per unit volume in a mixture of nanoparticles and small molecular weight solvent molecules can be estimated to be \footnote{ Assume that an arbitrary lattice of N sites is filled with  $N_1$  nanoparticles and  $N_2$  solvent molecules. Each nanoparticle incorporates x lattice sites and each solvent molecule one lattice site. Then the number of microstates  $\left(\Omega \right)\ $  is approximately $\ \Omega \approx N! /(\left(N-N_1\right)!N_1!)$ . Notice that  $N-N_1\ne N_2$  in contrast to the mixing theory of small molecules of same size. The Eq. (23) is obtained from the definition of entropy  $S=k_{{\rm B}}{\ln  \Omega \ }$  by simple algebraic manipulation and by assuming that the density of the nanoparticles is low ($N_1\ll N$ ).}
\begin{equation}
-\frac{TS}{V}=-\frac{k_{{\rm B}}T}{V_{{\rm S}}}\left[{\ln  \left(\frac{x}{x-\phi}\right)+\frac{\phi}{x}{\ln  \left(\frac{x-\phi}{\phi}\right)\ }\ }\right]\approx -\frac{k_{{\rm B}}T}{V_{{\rm S}}}\frac{\phi}{x}{\ln  \frac{x}{\phi}\ },
\label{23}
\end{equation}

where $V_{{\rm S}}$ is the volume of the solvent molecule, $\phi$ is the volume fraction of the nanoparticles and \textit{x} is the volume ratio between a nanoparticle and a solvent molecule. In the case of $x=1$ the equation properly reduces to
\begin{equation}
-\frac{TS}{V}=-\frac{k_{{\rm B}}T}{V_{{\rm S}}}\left[-\phi{\ln  \phi }-\left(1-\phi\right)\ln  \left(1-\phi\right)\right],\label{24}
\end{equation}
which corresponds to the entropy of mixing between two molecules of the same size.

For example, the volume of a toluene molecule is approximately $0.177\ {{\rm nm}}^{{\rm 3}}$ and the volume of a 10 nm nanoparticle is $524\ {{\rm nm}}^{{\rm 3}}$. In that case $x\approx 3000$. Eq. (\ref{23}) states that the entropy of mixing is reduced by a factor 1/1300 in a 1\% nanocomposite when compared to a situation in which both the nanoparticles and the solvent molecules were of the same size. Without a proof, it is suggested that the magnitude of the entropy is even less when the nanoparticles are mixed with polymer molecules. The suggestion is justifiable due to the entropic restrictions introduced by covalent bonding between the monomer units.

If the nanoparticles are magnetic, they interact with each other more strongly than non-magnetic nanoparticles. The magnetic dipolar interaction energy $\left(U_{{\rm M}}\right)$ between two particles, $2r$ in diameter, is given by \cite{chikazumi}
\begin{equation}
U_{{\rm M}}=\frac{{\mu }_0}{4\pi {\left(d+2r\right)}^3}\left(3\left({{\mathbf m}}_1\cdot \widehat{{\mathbf r}}\right)\left({{\mathbf m}}_2\cdot \widehat{{\mathbf r}}\right)-{{\mathbf m}}_1\cdot {{\mathbf m}}_2\right), \label{25}
\end{equation}
where $\widehat{{\mathbf r}}$\textbf{ }is the unit vector between the particles, \textit{d }is the distance between the particle surfaces and ${{\mathbf m}}_1$ and ${{\mathbf m}}_2$ are the magnetic moments of the particles. Assuming that the particles are magnetically single-domain, their saturation magnetization is $M_{{\rm S}}$ and that the magnetization vectors are parallel to each other and to the unit vector, the interaction energy is reduced to
\begin{equation}
U_{{\rm M}}=-\frac{8\pi }{9}\frac{r^6}{{\left(d+2r\right)}^3}{\mu }_0M^2_{{\rm S}}.\label{26}
\end{equation}
Similarly to the effective distance of the London dispersion force, one can derive the distance at which the magnetic energy is comparable to the thermal energy. It is given by
\begin{equation}
d_{k_{{\rm B}}T,{\rm MAGNETIC}}={\left(\frac{8\pi }{9}\frac{{\mu }_0M^2_{{\rm S}}}{k_{{\rm B}}T}\right)}^{\frac{1}{3}}r^2-2r.\label{27}
\end{equation}

To give an example, the magnetic interaction energy Eq. (\ref{26}) is drawn for two pairs of cobalt particles, 5 nm and 15 nm in diameter, in Figures \ref{figure2}A and \ref{figure2}B, respectively. The interaction between the 5 nm particles is dominated by the London dispersion potential and only weakly modified by the magnetic interaction. In the case of the 15 nm particles, the magnetic interaction is effective over a distance of 50 nm, rendering the London attraction negligible.
In order to shield magnetic nanoparticles from such a long-ranging interaction with a protective shell is unpractical. First of all, the maximum achievable nanoparticle volume fraction $\left({\hat{\phi}}_{{\rm MAGNETIC}}\right)$ is limited by the shielding. If the shielding layer volume is not taken to be a part of the nanoparticle volume, the maximum achievable volume fraction (neglecting entropic considerations) is proportional to
\begin{equation}
{\hat{\phi}}_{{\rm MAGNETIC}}\ \propto \frac{r^3}{{\left(d_{k_{{\rm B}}T,{\rm MAGNETIC}}+2r\right)}^3}\propto r^{-3}.\label{28}
\end{equation}
On the other hand, the maximum volume fraction $\left({\hat{\phi}}_{{\rm LONDON}}\right)$ limited by shielding against the London attraction does not depend on the nanoparticle size:
\begin{equation}
{\hat{\phi}}_{{\rm LONDON}}\propto \frac{r^3}{{\left(d_{k_{{\rm B}}T,{\rm LONDON}}+2r\right)}^3}={\rm const}.\label{29}
\end{equation}
Second, the shielding against the magnetic dipolar attraction by using the conventional grafting techniques is difficult due to the enormous length required from the grafted chains.

Based on the considerations presented in this Section, it is unlikely that a uniform dispersion of magnetic nanoparticles of decent size can be achieved by using the conventional shielding strategy. The magnetic interaction starts to dominate the free energy when the magnetic nanoparticles are 10 nm in diameter or larger. Furthermore, the entropic contribution decreases approximately as $x^{-1}$ where \textit{x }is the volume of the nanoparticle relative to the volume of the solvent molecule. Hence, the dispersion dilemma needs to be approached from some other point of view than the conventional shielding strategy.

\subsubsection{Effect of the polymer size, architecture, and functionalization}

A general dispersion strategy proposed by Mackay {\it et al.} suggests that the quality of a nanoparticle dispersion is strongly enhanced if the radius of gyration of the polymer is larger than the average diameter of the nanoparticle \cite{mackay}.  The radius of gyration  $\left(R_{{\rm G}}\right)$  for a polymer molecule which is interacting neutrally with its surroundings is given by $R_{{\rm G}}\approx \sqrt{{\rm C}/{\rm 6}}\sqrt{{\rm N}}{\rm a}$  where N is the number of monomers,  $a$  is the length of a single monomer and C is the Flory ratio. For the polystyrene that for example we use the equation yields 16 nm for the radius of gyration ($C\approx 9.9$ ,  $N\approx 2400$  and  $a\approx 0.25\ {\rm nm}$ ). It is based on the assumption that small particles can be incorporated within polymer chains easily but large particles prevent chains from achieving their true bulk conformations. In other words, large particles stretch the polymer molecules and hence introduce an entropic penalty. Pomposo et \textit{al}. have verified the Mackay's proposition in a material consisting of polystyrene and crosslinked polystyrene nanoparticles \cite{pomposo}. Such a system is ideal in a sense that the interaction between the polymer matrix and the nanoparticles is approximately neutral. That emphasizes the entropic contribution to the free energy. However, if the main contribution to the free energy is enthalpic, as it is in magnetic nanocomposites, one should use the Mackay's proposition with a considerable care. The entropic enhancement is most likely much smaller than the enthalpic term, rendering the improvement in the dispersion quality negligible.

One other remedy for the dispersion dilemma is to replace the linear polymer by a star-shaped one. It has been shown both theoretically \cite{balazs} and experimentally \cite{barnes} that it can lead to a spontaneous exfoliation of a polymer-nanoclay composite. It has been also demonstrated that replacing polystyrene in a polystyrene-nanoclay composite by a telechelic hydroxyl-terminated polystyrene results in exfoliation. Since the polymer-nanoclay composites are geometrically different from the polymer-nanoparticle composites, one cannot directly state that these techniques would also work with polymer-nanoparticles composites.

\subsection{Effective Magnetic Response}
\label{rules}

\begin{figure}[htb]
\includegraphics[width=7cm]{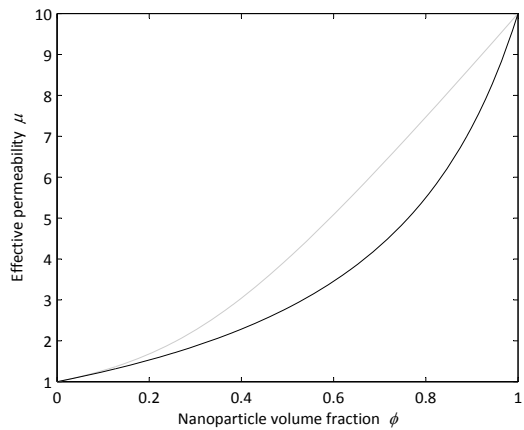}
\caption{The effective relative permeability of a nanocomposite containing spherical magnetic inclusions ($\mu _{\rm NP}$=10) as a function of the nanoparticle volume fraction. The Maxwell-Garnett theory prediction (black line) was obtained from Eq. (\ref{30}) and the Bruggeman theory prediction (grey line) from Eq. (\ref{31}).}
\label{fig3}
\end{figure}

The effective relative permeability of a nanocomposite containing spherical magnetic inclusions can be determined from several different effective medium theories (EMT) \cite{sihvola}. The two most popular are the Maxwell-Garnett formula
\begin{equation}
\mu =1+3\phi\frac{{\mu }_{{\rm NP}}{\rm -}1}{{\mu }_{{\rm NP}}+2-\phi\left({\mu }_{{\rm NP}}-1\right)},\label{30}
\end{equation}
and the symmetric Bruggeman formula
\begin{equation}
\frac{{\mu }_{{\rm NP}}-\mu }{{\mu }_{{\rm NP}}+2\mu }\phi +\frac{1-\mu }{1+2\mu }\left(1-\phi \right)=0,
\label{31}
\end{equation}
where $\mu $ is the effective relative permeability, ${\mu }_{{\rm NP}}$ is the relative permeability of the nanoparticles and $\phi$ is the nanoparticle volume fraction. The effective relative permeability of a nanocomposite containing spherical particles $\left({\mu }_{{\rm NP}}{\rm =10}\right)$ is plotted in Fig. \ref{fig3} according to both Eqs. (\ref{30}) and (\ref{31}). Below 20\% filling, the dependence of the permeability on the volume fraction is approximately linear. However, the rate of the linear increase is not as high as would be expected for homogeneous mixing.  The Bruggeman theory has been shown to agree with the experiments with similar materials as studied in this article \cite{paterson}. Before using the Bruggeman theory one needs to know what is the permeability of the nanoparticles. For uniaxial single-grain particles it is (\cite{chikazumi}, p. 439)

\begin{equation}
{\mu }_{{\rm NP},{\rm UNIAXIAL}}=1+\frac{{{\mu}_0M}^{{\rm 2}}_{{\rm S}}{{\sin }^{{\rm 2}} \theta \ }}{{\rm 2}K}\label{32}
\end{equation}
and for cubic particles

\begin{equation}
{\mu }_{{\rm NP},{\rm CUBIC}}=\left\{ \begin{array}{c}
1+\frac{{\mu }_0M^{2}_{{\rm S}}{{\sin }^{2} \theta \ }}{2K},\ \ K>0 \\
\\
1-\frac{{3{\mu }_0M}^{2}_{{\rm S}}{{\sin }^{2} \theta \ }}{4K},\ \ K<0 \end{array}
\right..
\label{33}
\end{equation}
where $\theta $ is the angle between the easy axis and the external field. Permeabilities of some ferromagnetic metals are calculated in Table \ref{table3}. It should be pointed out that once again the surface anisotropy has been neglected, and that it is most likely that the experimentally determined permeabilities are smaller than those in Table \ref{table3}.

\begin{table}
\center
\caption{The saturation magnetization $M_{\rm S}$ \cite{sorensen}, the anisotropy energy density (\textit{K}) \cite{sorensen,steinmuller}, and the calculated relative permeabilities $\mu_{\rm NP}$ (from Eq.(\ref{32}) and Eq.(\ref{33})) for selected ferromagnetic metals. $\langle\mu_{\rm NP}\rangle$ refers to the calculation where the permeability has been averaged over the isotropic distribution of the easy axes.}

\label{tabel}
\begin{tabular}{||c|c|c|c|c|}\hline\hline
    & $M_{S}$ & $K$ & $\mu_{\rm NP}$ & $\langle\mu_{\rm NP}\rangle$\\
   & (emu/cm$^3$) & (erg/cm$^3$) & ($\theta =\pi /2$) & \\
    \hline
   &  &  &  & \\
Iron (BCC) & 1707 & $4.8\times 10^5$ &  39& 26 \\
\hline
 &  &  &  &\\
Cobalt (HCP) & 1440 & $4.5\times 10^6$ &  4 &3  \\
\hline
&  &  &  & \\
Nickel (FCC) & 485 & $-5.7\times 10^4$ &  40 & 27
 \\
\hline
\hline

\end{tabular}
\label{table3}
\end{table}

The effective magnetic response of a polymer nanocomposite containing single-domain nanoparticles can be determined from the following rules:
\begin{itemize}
\item The FMR frequency determines the high-frequency limit of the magnetic response. The FMR frequency is determined from the effective anisotropy field by using Eq. (\ref{ll}).
\item The permeability below the FMR is dispersion-free since the only magnetization process taking place in single-domain particles is the domain rotation (which is associated with the FMR).
\item The magnitude of the permeability is determined from the Bruggeman theory, Eq. (\ref{31}).
\item The permeability of the nanoparticles - which is used in the Bruggeman theory - is determined from the effective anisotropy energy density and the saturation magnetization according to the Eqs. (\ref{32}) and (\ref{33}).
\end{itemize}

The anisotropy used in the calculations should be the true total anisotropy: the sum of the (bulk) magnetocrystalline anisotropy, the surface anisotropy, and the anisotropy due to magnetic  field. Especially if the bulk anisotropy is small, the surface anisotropy can be the dominant term. Since the experimental data on the surface anisotropy is scarce, it has been neglected in the analysis so far.

\section{Preparation and characterization}
\label{ch}

In this section we describe the experimental details and procedures used to prepare and characterize the nanocomposites.

\subsection{High volume fraction nanocomposites for high-frequencies}

Nanocomposites containing iron nanoparticles for the SHF band characterization were made according to the following procedure. First, a desired amount of nanoparticles (provided by Quantum Sphere, from now on abbreviated QS) were weighted and mixed with 15 ml of toluene (Fluka, purity better than 99.7\%). The desired amount of polystyrene was added and allowed to dissolve before vigurously sonicating the solution to break nanoparticle aggregates. The toluene was allowed to evaporate, resulting in a dark polystyrene-like film which was then collected.

Using this method, we have prepared nanocomposites of iron nanoparticles, with three different concentrations, 5\%, 10\% and 15\% (see Table \ref{table4}).

\begin{table}
\caption{Compositions of the nanocomposites prepared for electromagnetic characterization. }
\begin{tabular}{|c|c|c|}\hline
Designation & Nanoparticle type & $\varphi$ (\%) \\
    \hline\hline
PS/QS-Fe 5\% & Quantum Sphere Iron & 5 \\
\hline
PS/QS-Fe 10\% & Quantum Sphere Iron & 10 \\
\hline
PS/QS-Fe 15\% & Quantum Sphere Iron & 14.7 \\
\hline
\end{tabular}
\label{table4}
\end{table}

\subsection{Transmission electron microscopy: structural analysis}

\begin{figure}[htb]
\includegraphics[width=9cm]{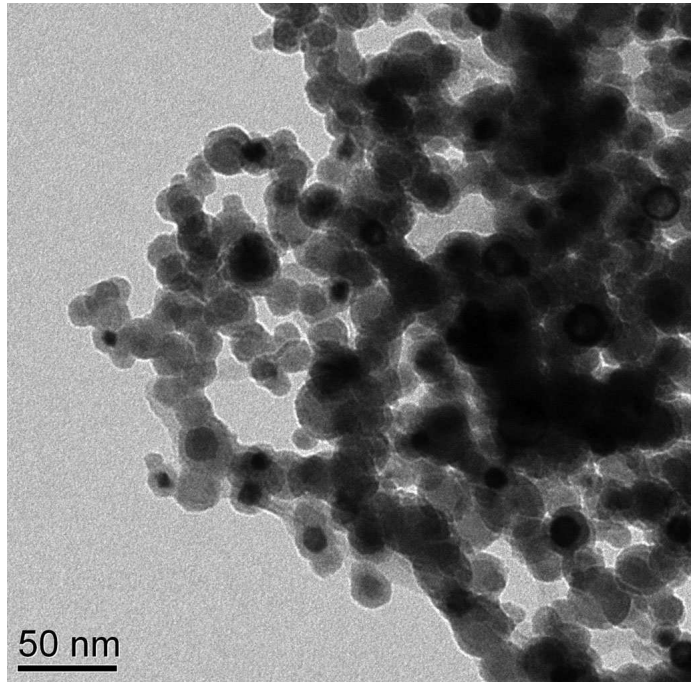}
\caption{Bright field TEM images of the  Quantum Sphere iron (scale bar is 50 nm). }
\label{fig4}
\end{figure}

The nanoparticles were imaged with a TEM (FEI Company model Tecnai G2 BioTwin) in bright field at the acceleration voltage of 120 kV. Before imaging the alignment of the microscope was checked and corrected. The image was recorded with a digital camera (Gatan model UltrascanTM 1000) and its contrast and brightness was adjusted after acquisition. An image of iron nanoparticles is shown in Figure \ref{fig4}.

\subsection{Magnetometry: low-frequency permeability}

Static hysteresis loops (the magnetization versus the applied field) of the nanoparticles were measured with a Superconducting Quantum Interference Device (SQUID) magnetometer (Quantum Design model MPMS XL) at 300 K. Roughly 1 mg of the nanoparticles was encapsulated in a piece of aluminum foil (approximately 100 mg) and attached to the plastic straw sample holder with Kapton tape. The permeability was extracted from the measured magnetization curve by fitting a straight line to the low-field part of the curve. The demagnetizing factor was approximated to be zero because the nanoparticles were compressed into flat layers inside the aluminum wrap and the layer surface was aligned along the external field.

\subsection{X-Ray Diffraction: Structure of the Nanoparticles}

\begin{figure}[htb]
\includegraphics[width=19cm]{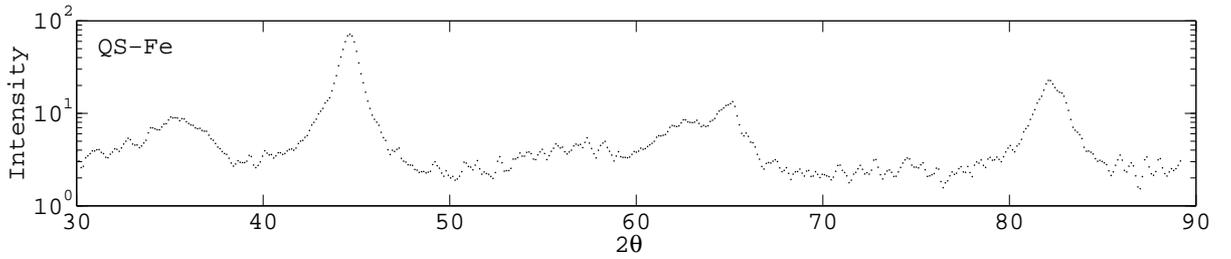}
\caption{XRD spectrum of iron nanoparticles.}
\label{xrd}
\end{figure}

The nanoparticle structure was analyzed with XRD. The diffraction intensities of the nanoparticles were measured as a function of the diffraction angle $2\theta$ with a diffractometer (PANalytical model X'Pert PRO MRD) using Cu K$_{\alpha}$ radiation (wavelength of 0.154056 nm) at room temperature. The XRD patterns of QS-Fe nanoparticles is shown in Figure \ref{xrd}. The samples were prepared by filling a circular cavity (35 mm in diameter and 0.7 mm high) bored into an acrylic glass plate with the particles. The powder was compressed and smoothed with a piece of a silicon wafer. The adhesion between the powder and the plate was sufficient to hold the powder within the cavity even though the plate was turned vertically for the measurement. The sample was scanned from 30$^\circ$ to 90$^\circ$ for one hour. The resulting data was processed by first stripping off the peaks due to the Cu K$_{\alpha_2}$ radiation and by filtering the background noise. The data was smoothed if the signal-to-noise ratio was poor. Second, the Lorentzian function was fitted to all peaks using the (self-implemented) Gauss-Newton algorithm. The performance of the algorithm was excellent in the case of well-defined peaks, but vague peaks had to be fitted manually. From the fitted peaks the angle, the FWHM and the intensity (integrated over the peak area) were extracted. Based on these values, the composition of nanoparticles was determined. Furthermore, the coherently scattering domain size was estimated from broadening of the FWHM. The natural width of a peak due to diffractometer was determined by measuring an annealed silicon powder sample and assuming that the coherently scattering domains were so large that their contribution to the broadening of the FWHM was negligible. The broadening due to the lattice strain was assumed to be minimal.

\subsection{Summary}

To summarize our results (see Table \ref{table5}), we find that The Quantum Sphere iron (QS-Fe) nanoparticles
are roughly 20-30 nm in diameter and most of the particles exhibit a core-shell structure.
Based on the XRD analysis, the core is suggested to comprise 9 nm BCC iron crystallites and the shell 3 nm FeO crystallites.
The composition was determined to be a 50\%-50\% balance between the oxide and the metal phases. The measured saturation magnetization (125 emu/g) is in rough agreement with the metal volume fraction estimated from XRD and the saturation magnetization given in literature for pure iron (218 emu/g) \cite{sorensen}.

\begin{table}
\caption{Summary of the nanoparticles and their properties. Particle diameters $d$ were estimated from the TEM images and the crystalline composition and the average crystallite diameters $d_{\rm CRYST}$ from the XRD measurement.}
\begin{tabular}{||c|c|c|c|c|c|c||}\hline\hline
    & $d$ & $M_S$ & $\mu$ & Crystal & $\varphi$ & $d_{\rm cryst}$ \\
   & nm & (emu/g) &  & (nm) & (\%) & (nm) \\
    \hline
Quantum Sphere Iron & 20-30 & 125 &  12.3 & Fe (BCC) & $50\pm 5$ & $9\pm 1$ \\
& &  &   & FeO & $50\pm 5$ & $3\pm 1$ \\
\hline
\hline
\end{tabular}
\label{table5}
\end{table}

\section{High-frequency properties}
\label{hf}

\subsection{Coaxial airline technique: permittivity and permeability in the SHF band}
\label{prop}

A broadband coaxial airline method  developed in \cite{chalapat} was used to measure the complex
permittivity and the complex permeability of magnetic composites in the
superhigh-frequency band (SHF). The technique involves measurement of the reflection
parameters $S_ {11}$ and $S_{22}$, the transmission parameters $S_{12}$ and $S_{21}$, and the group delay through a sample inserted inside a 7 mm precision coaxial airline. The measurement was done by
connecting the coaxial airline to a vector network analyzer (Rohde and Schwarz ZVA40)
using a pair of high-performance cables (Anritsu 3671K50-1). Prior to the
measurement, the errors due to the loss and reflection in the cables, connectors and the
network analyzer were removed by performing a SOLT calibration up to both ends of the RF
cables.

The sample required in the coaxial waveguide measurement is a cylinder, 7.00 mm in diameter, with a 3.04 mm hole in the middle. Its thickness can be adjusted between 4 mm and 10 mm in order to avoid the dimensional resonance. The samples were made by hot-pressing each nanocomposite inside a polished 7 mm hole drilled through a steel plate. Prior to the pressing, the plate and the nanocomposites inside the holes were sandwiched between two sheets of poly(ethylene terephthalate) and further between two solid steel plates.  The assembly was inserted into a hot-press (Fontijne model TP 400) at 160 ${}^\circ$C and kept there for two minutes. After the nanocomposite had softened, a 400 kN force was applied over the plates. After waiting for another two minutes, the pressure was released and the plate system was disassembled. The holes containing the softened and compressed nanocomposites were refilled and the pressing was done again. The filling was repeated one more time. After the third pressing, the assembly was placed between two metal plates cooled with circulating water under a 400 kN force.  After the plates had cooled down to room temperature, the pressure was released and the plates were disassembled. Before detaching the solidified cylindrical samples, the plate with the samples was sandwiched between two 5 mm thick steel plates with 3.1 mm holes exactly above and under of each of the 7 mm holes in the central plate. All the three plates were aligned with respect to each other and clamped together. The assembly was fixed under a vertical boring machine and holes were drilled through the nanocomposites through the guiding 3.10 mm holes in the upper plate. The used drill bit was 2.9 mm in diameter since it was found out that the drilling produced holes 0.1---0.2 mm wider than the drill bit. After drilling, the construction was disassembled and the samples were detached by gently pushing them out of the holes. If necessary, the pellets were finished by carefully removing any imperfections with sandpaper. The sample dimensions were measured with a caliper.

\begin{figure}[h!]
 \centering
 \includegraphics[scale=0.35]{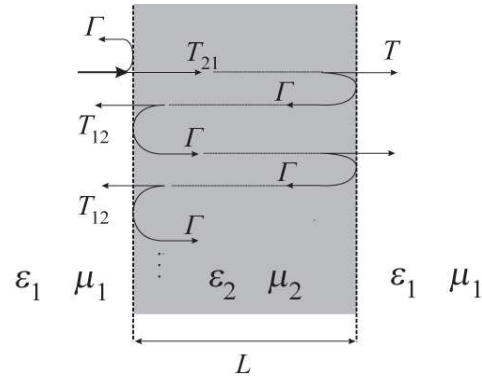}
 \caption{The model of multiple reflection between two interfaces. Figure republished with permission (\copyright IEEE 2009) from \cite{chalapat}.}
 \label{MultipleReflection}
\end{figure}

Below we briefly present the measurement method; the mathematical relations between the
S-parameters and the material parameters are given. More discussions, including detailed
error analysis of the method we use, are presented in \cite{chalapat}. Basically, the method is
developed based on the multiple reflection model, as shown schematically in Fig.
\ref{MultipleReflection}.

When the wave arrives at the first interface at $z=0$, the reflection and transmission
occurs. This means part of the wave is reflected with a coefficient $\Gamma$, and part of it is
transmitted with a coefficient $T_{21}$. The transmitted wave then travels through the
second medium and gets reflected again at the second interface with a coefficient $\Gamma$
while part of it is transmitted through the second interface with a coefficient $T_{12}$. It
can be seen from Fig. \ref{MultipleReflection} that this transmission and reflection
continuously occurs (ideally) an infinite number of times or until the wave has lost all of its
energy.

To find the total reflection coefficient in this model, we need to sum up all the
reflected waves. The superposition of waves can be calculated in the same way as the
summation of vectors in which both amplitude
and phase must be considered. We know that a wave traveling a distance $L$ through the
second medium has a propagation factor given by
\begin{equation}
P = e^{-\gamma_2 L},
\label{P}
\end{equation}
where $\gamma_2 = i\omega/v_2 = i\omega n_2/c$.

The total reflection coefficient can then be expressed as follows
\begin{equation}
\Gamma_{\rm tot} = \Gamma + T_{21}T_{12}\Gamma P^2 + T_{21}T_{12}\Gamma^3 P^4 + ...
= \frac{\Gamma (1-P^2)}{1- \Gamma^2 P^2},\nonumber
\label{Gamma_1}
\end{equation}
where
\begin{equation}
\Gamma = \frac{1-\sqrt{\frac{\epsilon_2 \mu_1}{\epsilon_1 \mu_2}
}}{1+\sqrt{\frac{\epsilon_2 \mu_1}{\epsilon_1 \mu_2} 
}},
\label{GammaDef}
\end{equation}
and
\begin{equation}
T_{12} = 1 + \Gamma = \frac{2}{1+\sqrt{\frac{\epsilon_1 \mu_2}{\epsilon_2 \mu_1} }} =
\sqrt{\frac{\epsilon_2 
\mu_1}{\epsilon_1 \mu_2}} \hspace{2pt}T_{21}. \label{T12}
\end{equation}

Similarly, the total transmission coefficient in terms of $\Gamma$ and $P$ is
\begin{equation}
T_{\rm
tot} = \frac{P \left(1 - \Gamma^2\right)}{1-\Gamma^2 P^2}.
\label{Teq4}
\end{equation}

\begin{figure}[h!]
 \centering
 \includegraphics[scale=0.4]{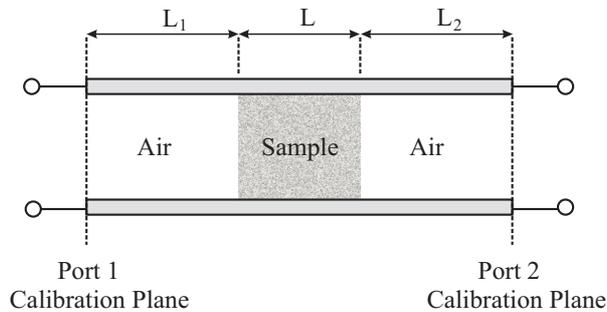}
 \caption{The diagram of a transmission line containing two interfaces and the planes at
which scattering parameters are measured. Figure republished with permission (\copyright 2009 IEEE)
from \cite{chalapat}.}
 \label{TransmissionReflection}
\end{figure}

In practice, the study of discontinuities within a transmission line is done via the
measurement of S-parameters. Considering the measurement setup as illustrated
schematically in Fig. \ref{TransmissionReflection}, we can see that when a wave travels
from the first port to the first interface, it accumulates a phase change of $-\gamma_1
L_1$, where $\gamma_1 = i\omega n_1/c \approx i\omega/c$. Similarly, from the second
interface to the second port, it will pick up another phase change of $-\gamma_1 L_2$.
This means
\begin{equation}
S_{21} = S_{12} = e^{-\gamma_1 (L_1 + L_2)} T_{tot} = e^{-\gamma_1 (L_1 + L_2)}\frac{P(1
- \Gamma^2)}{1-\Gamma^2P^2},
\label{S21}
\end{equation}
\begin{equation}
S_{11} = e^{-2 \gamma_1 L_1} \Gamma_{tot} = e^{-2 \gamma_1 L_1} \frac{ \Gamma
\left(1-P^2\right)}{1-\Gamma^2P^2},
\label{S11}
\end{equation}
and
\begin{equation}
S_{22} = e^{-2 \gamma_1 L_2}\Gamma_{tot} = e^{-2 \gamma_1 L_2} \frac{ \Gamma
\left(1-P^2\right)}{1-\Gamma^2P^2}.
\label{S22}
\end{equation}
In principle, there are many ways to obtain the material parameters based on the above
equations. The method presented here is chosen because it does not require the
measurement of $L_1$ and $L_2$; as a result, material parameters can be accurately determined.

The algorithm proceeds further by first defining two reference-plane invariant
quantities, namely
\begin{equation}
A = \frac{S_{11}S_{22}}{S_{21}S_{12}} =
\frac{\Gamma^2}{(1-\Gamma^2)^2}\frac{(1-P^2)^2}{P^2},
\label{Adef}
\end{equation}
and
\begin{equation}
B = e^{2 \gamma_1 (L_{\rm air} - L)}(S_{21}S_{12}-S_{11}S_{22}) = \frac{P^2 -
\Gamma^2}{1- \Gamma^2 P^2}.
\label{Bdef}
\end{equation}
\noindent Next, Eq. (\ref{Bdef}) is solved for $P^2$,
\begin{equation}
P^2 = \frac{B +  \Gamma^2}{1 + B \Gamma^2}.
\label{P2BC}
\end{equation}

Then, simply by substituting $P^2$ into (\ref{Adef}), a new expression for $A$
is obtained,
\begin{equation}
A = \frac{\Gamma^2 (1 - B)^2}{(B + \Gamma^2)(1 + B \Gamma^2)},
\label{ABGamma}
\end{equation}
which gives us
\begin{equation}
\Gamma^2 =\frac{-A(1+B^2)+(1-B)^2}{2AB}
\pm \frac{\sqrt{-4A^2B^2+\left[A(1+B^2)
-(1-B)^2\right]^2}}{2AB}, \label{Gamma2AB}
\end{equation}
where the sign in this equation is chosen so that  $|\Gamma| \leq 1$. As we can see,
these expressions for $P^2$ and $\Gamma^2$ are reference-plane invariant.

In the next step, another quantity is defined, namely
\begin{equation}
R = \frac{S_{21}}{S_{21}^o} = \frac{e^{\gamma_1 L} P (1-\Gamma^2)}{1-P^2\Gamma^2}.
\label{Rdef}
\end{equation}

\noindent Substituting $P^2$ from Eq. (\ref{P2BC}) into Eq. (\ref{Rdef}), we get a new
expression for $P$,
\begin{equation}
P = R\frac{1+\Gamma^2}{1+B\Gamma^2}e^{-\gamma_{1}L}.
\label{PRGamma}
\end{equation}

By using Eqs. (\ref{Gamma2AB}) and (\ref{PRGamma}), we can determine the other
constitutive parameters of materials, for example the complex index of refraction,
\begin{equation}
n = n'+in'' = \sqrt{\mu_r \epsilon_r} = \frac{1}{\gamma_1 L}\ln\left(\frac{1}{P}\right).
\label{nn}
\end{equation}

Similar to the Nicolson-Ross Weir algorithm, \cite{NicolsonRoss}-\cite{Weir}, the method
requires the evaluation of group delay for choosing the correct result. But, it should be
noted that, only the real part of $n$, in Eq. (\ref{nn}), is multi-valued, the imaginary
part is not, {\it i.e.} every root provides the same $n''$. So measuring the imaginary part of
the index of refraction does not require the evaluation of the group delay. This concept
could be practically useful for examples when energy loss resonances are studied.

In case of non-magnetic materials, determining the complex permittivity from $\epsilon_r
= n^2$ provides a better alternative relative to the NRW method.
This is because, this way, one does not need to calculate the
relative wave impedance $z = (1+\Gamma)/(1-\Gamma)$, which exhibits high errors at and
around the Bragg resonance frequencies \cite{Boughriet}.

As discussed in \cite{chalapat}, extra steps must be done if this method is applied to
measure materials with unknown magnetic properties. One way to do so, is to simply use
one of the roots $\pm\Gamma$ of Eq. (\ref{Gamma2AB}), and simultaneously plot the spectra
of both $\epsilon_r$ and $\mu_r$. Then, based on chemical analysis, the permeability
spectra can be extracted. This algorithm is based on the fact that the sign of $\Gamma$
only swaps the values of permittivity and permeability.

\subsection{Nanocomposites: permittivity and permeability in the SHF band}
\label{ss}

\begin{figure}[htb]
\includegraphics[width=9cm]{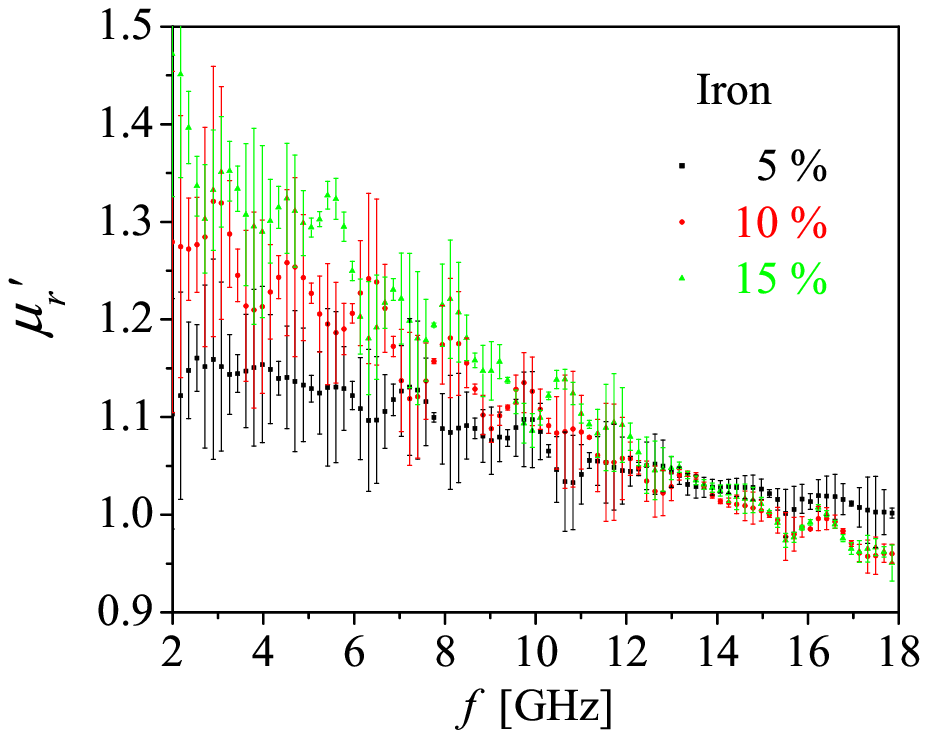}\includegraphics[width=9cm]{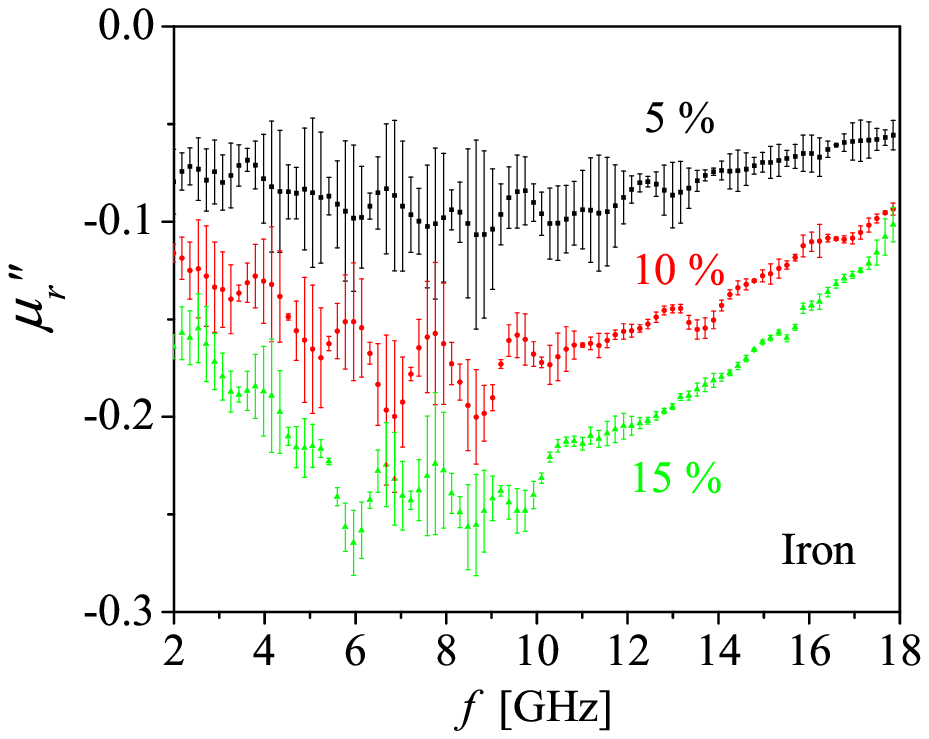}
\caption{The complex relative permeability (real part in the left figure and imaginary part in the right figure) of  PS/QS-Fe nanocomposites between 2
and 12 GHz. The black data points are from the 5\% sample, the red data points from
the 10\% sample and the green data points from the 15\% sample.}
\label{u}
\end{figure}

\begin{figure}[htb]
\includegraphics[width=9cm]{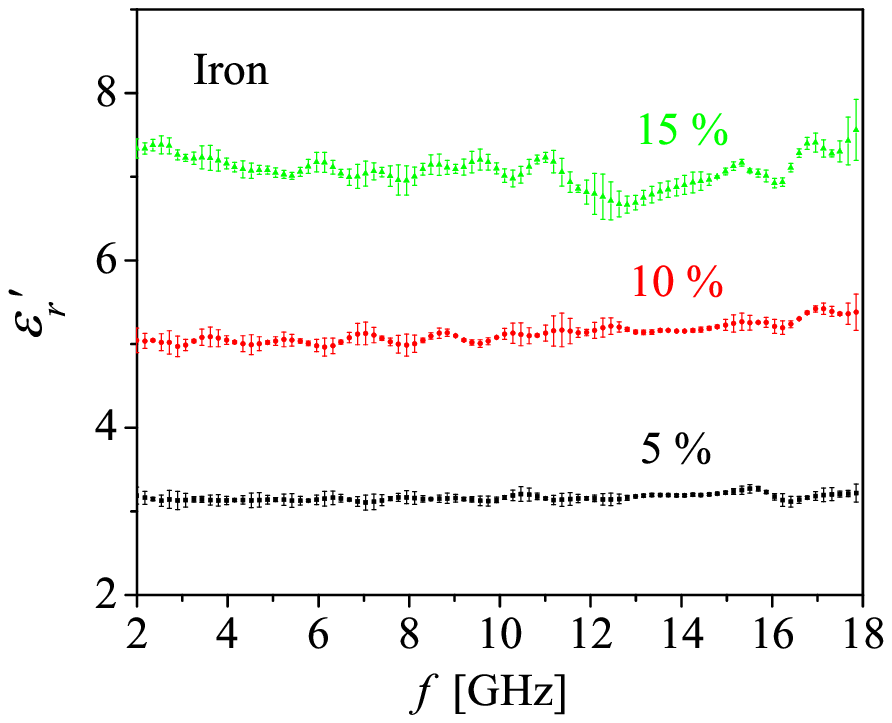}\includegraphics[width=9cm]{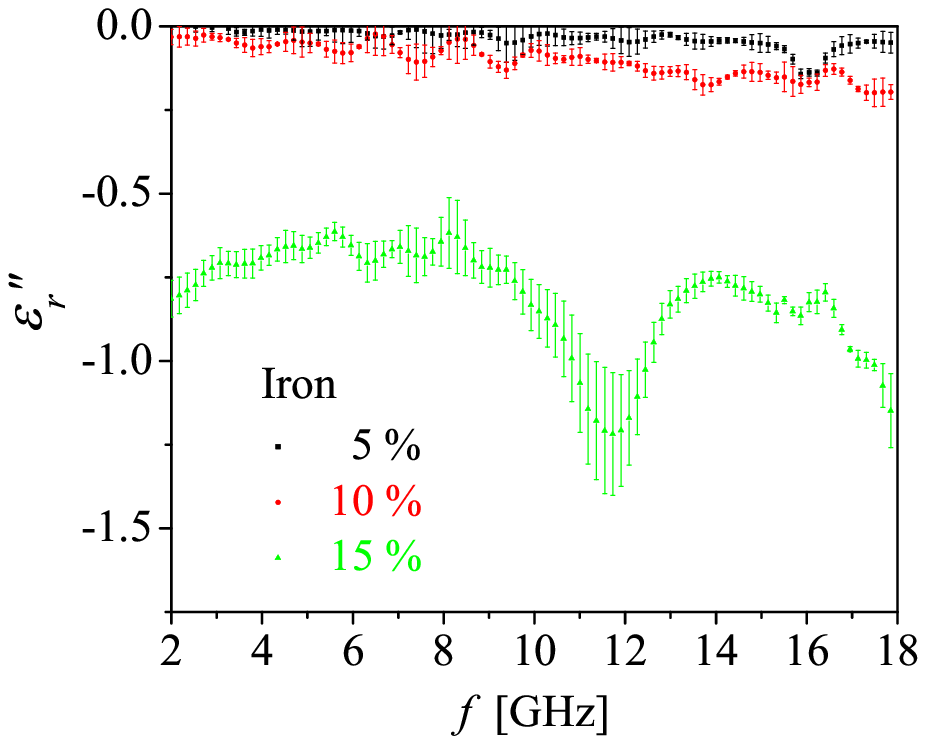}
\caption{The complex relative permittivity (real part in the left figure and imaginary part in the right figure) of  PS/QS-Fe nanocomposites between 2
and 12 GHz. The black data points are from the 5\% sample, the red data points from
the 10\% sample and the green data points from the 15\% sample.}
\label{e}
\end{figure}

\begin{figure}[htb]
\includegraphics[width=9cm]{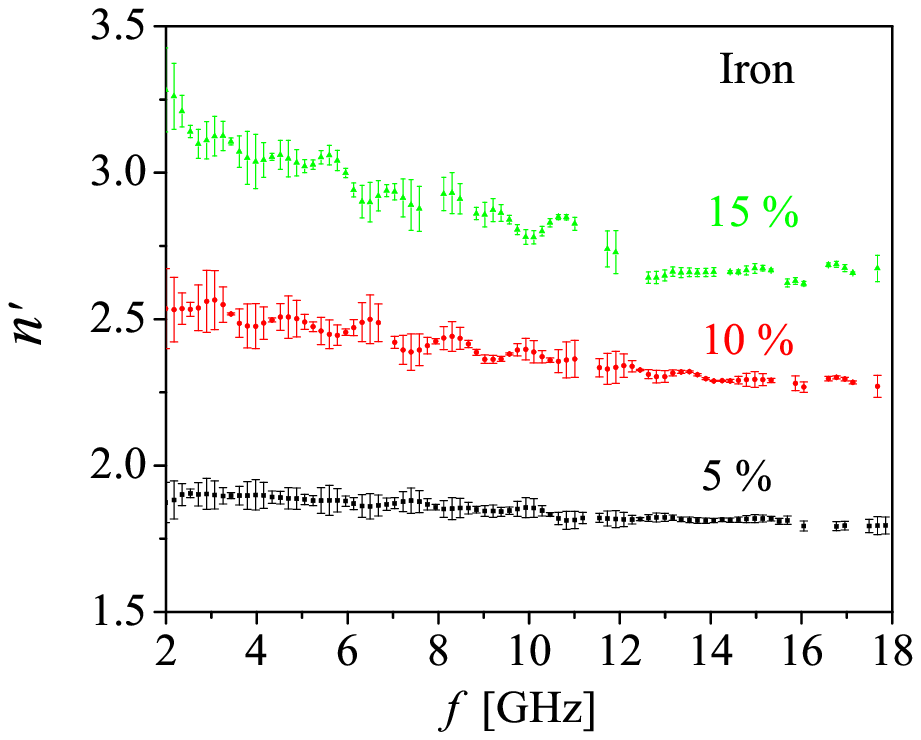}\includegraphics[width=9cm]{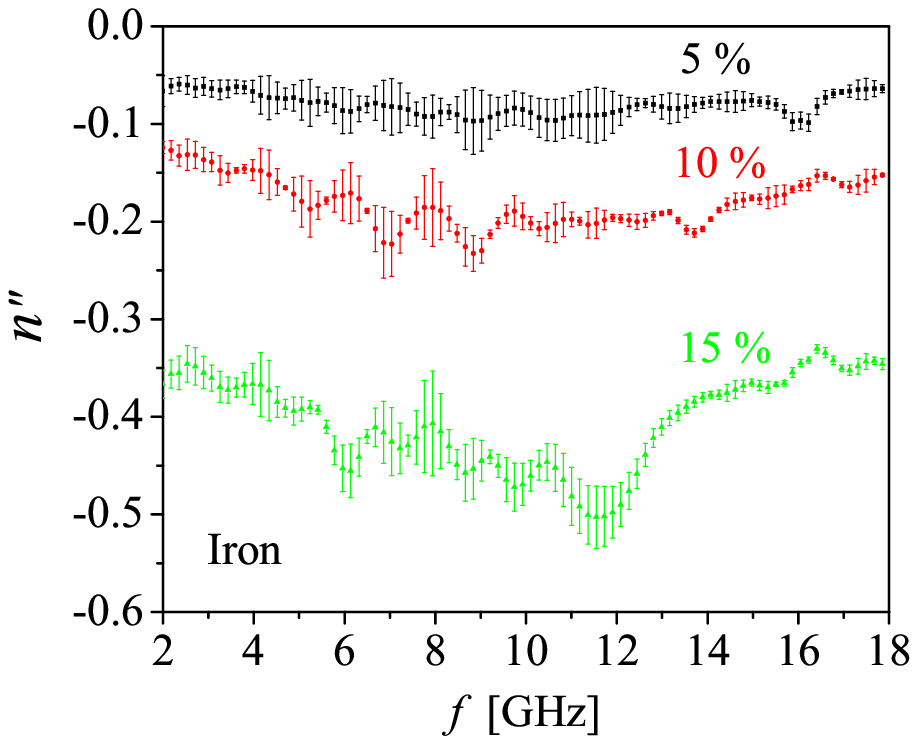}
\caption{The complex index of refraction (real part in the left figure and imaginary part in the right figure) of  PS/QS-Fe nanocomposites between 2
and 12 GHz. The black data points are from the 5\% sample, the red data points from
the 10\% sample and the green data points from the 15\% sample.}
\label{n}
\end{figure}

We now  present our experimental results corresponding to a nanocomposite comprising iron nanoparticles (20--30 nm, BCC)
in polystyrene (PS/QS-Fe) (Figures \ref{e},\ref{u},\ref{n}).

All the composites 5\%, 10\% and 15\% exhibited mild ferromagnetic resonances between 6 GHz and 8 GHz. These resonances correspond to the anisotropy fields between 0.20 T and 0.27 T. The expected anisotropy field calculated from bulk BCC iron
magnetocrystalline anisotropy is 56 mT. This large difference may be
explained by additional anisotropy components due to surface effects or
particle-particle interactions. In the case of surface anisotropy the
broadening of the resonance peak would be due to the finite size
distribution of the nanoparticles, and in the case of particle-particle
interactions due to variations of the polarizing field due to irregular
spatial arrangement and orientation of the particles. The Snoek limit (Eq. \ref{result}) predicts that no higher relative magnetic permeability than 8.5 can be achieved at 5 GHz in (positive) uniaxial and cubic materials which are either bulk or composites containing spherical inclusions. According to the Bruggeman theory (Eq. \ref{31}) the effective relative permeability is at maximum one sixth of 8.5 in a nanocomposite containing less than
15\% magnetic inclusions. We find a relative permeability of the order $\mu =1.3$
in the 15\% nanocomposite, which is already pushing the Snoek limit.
The exact determination of whether the Snoek limit has been exceeded depends on the nature
of the anisotropy which would require precise knowledge of the surface contribution.

The permeabilities (both real and imaginary) in all composites were roughly
constants over the measurement range. While the real part increased roughly
linearly with the volume fraction, there was much larger jump in the
imaginary part from 10\% to 15\% than from 5\% to 10\%. This behavior could be
attributed to the increase in conductivity due to exceeding the percolation
threshold.

The relative permittivities were observed to increase with the nanoparticle volume fraction from approximately 2.5, which is a typical value for polystyrene. The imaginary parts were observed to be significantly larger than $10^{-4}$ (a typical value for pure polystyrene). The increase in both the real part and the imaginary part is understood to be due to the electrical polarizability of the nanoparticles in the electric field. The dispersion for the 5\% and the 10\% composites was also comparatively lower than for the 15\% sample (Figure \ref{n}).

\section{Conclusions: how to improve the performance in the SHF band}
\label{conc}

The high-frequency magnetic performance is always a compromise between the permeability and the FMR frequency. Increasing the magnetic anisotropy, no matter wherefrom it originates, decreases the permeability but increases the FMR frequency. Only increasing the saturation magnetization increases both the permeability and the FMR frequency (Eq. \ref{result}). Hence, the saturation magnetization should be maximized while a compromise needs to be done with the anisotropy. Further degree of freedom stems from the shape of the magnetic inclusions. It is well known that the resonance frequency of an arbitrary magnetic body with the demagnetization factors $N_x$, $N_y$ and $N_z$ is given by the Kittel formula \cite{kittel}
\begin{equation}
f_{{\rm FMR}}=(2\pi)^{-1} \nu \sqrt{\left[H_{{\rm A}}+\left(N_x-N_z\right)M_{{\rm S}}\right]\left[H_{{\rm A}}+\left(N_y-N_z\right)M_{{\rm S}}\right]}, \label{43}
\end{equation}
from which the well known resonance formulas for bulk, film, rod and sphere can be derived ($\nu$ is the gyromagnetic constant, defined as in Eq. (\ref{nu})). In the cases of spheres ($N_{x}=N_{y}=N_{z}=1/3$) and bulk material ($N_{x}=N_{y}=N_{z}=0$), the resonance frequency is directly proportional to the anisotropy field (as was assumed in Subsection \ref{snoek}). In the infinite rod limit ($N_{x}=N_{y}=1/2$, $N_{z}=0$) the FMR frequency is linearly proportional to the saturation magnetization (assuming that $M_{{\rm S}}\gg H_{{\rm A}}$) and in the thin film limit ($N_{x}=N_{y}=0$, $N_{z}=1$) to the square root of the saturation magnetization and the anisotropy field (assuming that $M_{{\rm S}}\gg H_{{\rm A}}$). Even in the case of HCP cobalt, which has a high anisotropy field of ${{\mu }_0H}_{{\rm A}}\approx 0.63\ {\rm T}$, the highest resonance frequency is obtained in the non-isotropic geometries, namely the infinite rod and the thin film. The same conclusion is valid for the BCC iron (${{\mu }_0H}_{{\rm A}}\approx 56\ {\rm mT})$, and the FCC nickel (${\mu }_0H_{{\rm A}}\approx 16\ {\rm mT})$. However, due to the surface anisotropy,  in the end the anisotropy field in nanoscale rods, spheres and films can be much larger than in bulk. It should be understood that the FMR frequency depends on the shape of the magnetic inclusions, but obtaining qualitative results from calculations without knowing the surface anisotropy is not possible (as already pointed out in Subsection \ref{snoek}). In addition, the FMR frequency depends on particle-particle interactions.

The volume fraction of the inclusions has obviously an effect on both the resonance frequency and the permeability. The permeability of a nanocomposite with spherical inclusions can be calculated directly from the Bruggeman theory (Subsection \ref{rules}) but in all the other cases the Bruggeman equation must be solved iteratively and self-consistently with the Landau-Lifshitz equation \cite{ramprasad}. The results from such calculations indicate that 1) the FMR frequency of a nanocomposite containing spherical inclusion does not depend on the volume fraction and 2) in all other cases the FMR frequency smoothly varies from the single-inclusion limit to the homogeneous bulk limit. Hence, the set of rules for predicting the high-frequency magnetic performance stated in Subsection \ref{rules} are valid only for spherical nanoparticles. As argued above, the FMR resonance is the lowest in the bulk and spherical particle limits.  The results presented in Subsection \ref{ss} agree with the literature in a sense that the FMR frequency was found out to vary only a little with the nanoparticle volume fraction. The small variation might be due to slight deviations from ideal spherical form or due to the aggregation of the nanoparticles.

The magnetic performance in nanocomposites could be improved by taking all the above considerations into account when designing the material. In addition, it is preferential to use monodisperse, single-crystal nanoparticles in order to observe well-defined resonance peaks. Without such information, quantitative evaluation of the magnetic performance is difficult. Also, the surface effects such as the surface anisotropy have to be taken into account since nanoparticles have a huge surface-to-volume ratio compared to bulk materials. Because the magnitude and the symmetry of the surface anisotropy is difficult to calculate, it cannot be taken in practice into consideration before the measurement. Instead, it is the deviation of the observed FMR frequency from the value expected from bulk magnetocrystalline anisotropy that indicates the magnitude of the surface anisotropy.
After having decided the target permeability and resonance frequency and having approximated the type and the volume fraction of the magnetic inclusions, one still needs to find the appropriate processing route which can lead to such a nanocomposite. As argued in this article, and also accepted in literature, homogeneous blends of plain nanoparticles in polymers are almost impossible to achieve even at the lowest filling ratios. In the high-frequency applications the role of the nanoparticles is not just an additive since the practical volume fractions (from the application perspective) begin from 10\%. Hence, the nanoparticles should be a supporting part of the nanocomposite --- not an additive.

Suppressing the large permittivity and dielectric loss will be a difficult task. The imaginary losses can be reduced by using single-crystal nanoparticles in which the conduction electron scattering is suppressed. Tackling the real part is much more difficult, since all metallic nanoparticles are highly conductive and their polarizability should of the same order.

\section{Acknowledgements}
This work was supported by the Finnish Funding Agency for Technology and Innovation (TEKES).
G.S.P. would like to acknowledge also partial support from the Academy of Finland (Acad. Res. Fellowship 00857 and projects 129896, 118122, and 135135). K.C. wishes to thank the Thailand Commission on Higher Education for financial support.


\begin{thebibliography}{99}

\bibitem{shirakata} Y. Shirakata, {\it et al.}, "High Permeability and Low Loss Ni-Fe Composite Material for High-Frequency Applications", IEEE Transactions on Magnetics, {\bf 44}, 2100 (2008).

\bibitem{hansen} R. C. Hansen and M. Burke, "Antennas with magneto-dielectrics", Microwave and Optical Technology Letters {\bf 26}, 75 (2000).

\bibitem{waldron} R. A. Waldron, Ferrites: an introduction for microwave engineers, (Van Nostrand, London, 1961).

\bibitem{super} "Super high frequency", $http://en.wikipedia.org/wiki/Super\_high\_frequency$ (2.2.2009).

\bibitem{sun} S. Sun, {\it et al.}, "Monodisperse FePt Nanoparticles and Ferromagnetic FePt Nanocrystal Superlattices", Science {\bf 287}, 1989 (2000).

\bibitem{puntes} V. F. Puntes, {\it et al.}, "Colloidal Nanocrystal Shape and Size Control: The Case of Cobalt", Science {\bf 291}, 2115 (2001).

\bibitem{moerup} S. Moerup and M. F. Hansen, "Fundamentals and Theory", in Handbook of Magnetism and Advanced Magnetic Materials, edited by H. Kronmueller and S. Parkin, (John Wiley \& Sons, 2007).

\bibitem{wu} L. Z. Wu, {\it et al.}, "Studies of high-frequency magnetic permeability of rod-shapes CrO2 nanoparticles", Phys. Stat. Sol. {\bf 204}, 755 (2006).

\bibitem{desvaux} C. Desvaux, {\it et al.}, "Multimillimetre-large superlattices of air-stable iron-cobalt nanoparticles", Nat. Mater. {\bf 4}, 750 (2005).

\bibitem{chikazumi} S. Chikazumi, Physics of Ferromagnetism, (Oxford University Press, New York, 1997).

\bibitem{handley} R. C. O'Handley, Modern Magnetic Materials, (John Wiley \& Sons, Inc., 2000).

\bibitem{kittel} C. Kittel, Introduction to Solid State Physics, (John Wiley \& Sons, New York, 1971).

\bibitem{sorensen} C. M. Sorensen, Magnetism in Nanoscale Materials in Chemistry, edited by K. J. Klabunce, (Wiley-IEEE, 2001).

\bibitem{steinmuller} S. J. Steinmuller, {\it et al.}, "Effect of substrate roughness on the magnetic properties of thin fcc Co films", Physical Review B {\bf 76}, (2007).

\bibitem{george} P. K. George and E. D. Thompson, "Exchange Stiffness in Hexagonal Cobalt", Physical Review Letters {\bf 24}, 1431 (1970).

\bibitem{snoek} J. L. Snoek, "Dispersion and absorption in magnetic ferrites at frequencies above one Mc/s", Physica {\bf 15}, 207 (1948).

\bibitem{landau} L. Landau and E. Lifshitz, "On the theory of the dispersion of magnetic permeability in ferromagnetic bodies'", Phys. Z. Sowjetunion 8, (1935).

\bibitem{jonker} G. H. Jonker, {\it et al.}, Philips Tech. Rev. 18, 1956 (1956).

\bibitem{krishnamoorti} R. Krishnamoorti and R. A. Vaia, "Polymer Nanocomposites", Journal of Polymer Science: Part B: Polymer Physics {\bf 45}, 3252 (2007).

\bibitem{mackay} M. E. Mackay, {\it et al.}, "General Strategies for Nanoparticle Dispersion", Science {\bf 311}, 1740 (2006).

\bibitem{balazs} A. C. Balazs and S. Chandralekha, "Effect of polymer architecture on the miscibility of polymer/clay mixtures", Polymer international {\bf 49}, 469 (2000).

\bibitem{glogowski} E. Glogowski, {\it et al.}, "Functionalization of Nanoparticles for Dispersion in Polymers and Assembly in Fluids", Journal of Polymer Science: Part A: Polymer Chemistry {\bf 44}, 5076 (2006).

\bibitem{latham} A. H. Latham and M. E. Williams, "Controlling Transport and Chemical Functionality of Magnetic Nanoparticles", Acc. Chem. Res. {\bf 41}, 411 (2008).

\bibitem{isra} J. Israelachvili, "Intermolecular and Surface Forces", (Academic Press, London, 1985).

\bibitem{pomposo} J. A. Pomposo, {\it et al.}, "Key role of entropy in nanoparticle dispersion: polystyrene-nanoparticle/linear-polystyrene nanocomposites as a model system", Phys. Chem. Chem. Phys. {\bf 10}, 650 (2007).
	
\bibitem{barnes} C. Barnes, {\it et al.}, "Spontaneous Formation of an Exfoliated Polystyrene-Clay Nanocomposite Using a Star-Shaped Polymer", J. Am. Chem. Soc. {\bf 126}, 8118 (2004).

\bibitem{sihvola} Sihvola, "Electromagnetic mixing formulas and applications", (The Institution of Electrical Engineers, 1999).

\bibitem{paterson} J. H. Paterson, {\it et al.}, "Complex permeability of soft magnetic ferrite/polyester resin composites at frequencies above 1 MHz", Journal of Magnetism and Magnetic Materials {\bf 196-197}, 394 (1999).

\bibitem{klug} H. P. Klug and L. E. Alexander, X-Ray Diffraction Procedures, (John Wiley and Sons, 1954).

\bibitem{bjorck} \AA . Björck, Numerical methods for least squares problems, (SIAM, Philadelphia, 1996).

\bibitem{chalapat} K. Chalapat, K. Sarvala, J. Li, and G. S. Paraoanu,
"Wideband Reference-Plane Invariant Method for Measuring Electromagnetic Parameters of Materials",
IEEE Trans. Microw. Theory Tech.
{\bf 57}, 2257 (2009).

\bibitem{NicolsonRoss}
A.~M.~Nicolson and G.~F. Ross, "Measurement of the intrinsic properties of materials by time-domain 
techniques", IEEE Trans. Instrum. Meas., {\bf IM-19}, pp. 377-382, Nov. 1970.

\bibitem{Weir}
William~B.~Weir, "Automatic measurement of complex dielectric constant and permeability at microwave 
frequencies", Proc. IEEE, {\bf 62}, pp. 33-36, Jan. 1974.

\bibitem{Boughriet}
A.-H.~Boughriet,~C.~Legrand and A.~Chapoton, "Noniterative stable transmission/reflection method for low-loss material 
complex permittivity determination", IEEE Tran. Microw. Theory Tech., {\bf  45}, pp. 52-57, Jan. 1997.

\bibitem{ramprasad} R. Ramprasad {\it et al.}, "Fundamental Limits of Soft Magnetic Particle Composites for High Frequency Applications", Phys. Stat. Sol. {\bf 233}, 31 (2002).

\end{thebibliography}
\end{document}